\documentclass[article,12pt]{elsarticle}
\usepackage{amsmath}
\usepackage{float}
\usepackage{comment}

\usepackage{amssymb}

\journal{Applied Mathematics and Computation}

\begin{document}
\begin{frontmatter}


\title{Spectral problem for a two-component nonlinear Schr\"odinger equation in $2+1$ dimensions: Singular manifold method and Lie point symmetries}

\author{P. Albares$^{a)}$, J. M. Conde$^{b)}$  and P. G. Est\'evez$^{a)}$}

\address{$^{a)}$ Departamento de F\'{\i}sica Fundamental, Universidad de Salamanca, Salamanca, Spain}

\address{$^{b)}$ Universidad San Francisco de Quito (USFQ), Quito, Ecuador.
Departamento de Matem\'aticas, Colegio de Ciencias e Ingenier\'ias}

\begin{abstract}

An integrable  two-component nonlinear Schr\"odinger equation in $2+1$ dimensions is presented. The singular manifold method is applied in order to obtain a three-component Lax pair. The Lie point symmetries of this Lax pair are calculated in terms of nine arbitrary functions and one arbitrary constant that yield a non-trivial infinite-dimensional Lie algebra. The main non-trivial similarity reductions associated to these symmetries are identified. The spectral parameter of the reduced spectral problem appears as a 
consequence of one of the symmetries.
\end{abstract}

\end{frontmatter}
\section{Introduction}
Invariance of a differential equation under a set of transformations is equivalent to the existence of symmetries. The study of symmetries represents a fundamental aspect related to the analysis of integrability of dif\-fer\-en\-tial equations, since this invariance property may be used to achieve partial or complete integration of such equations \cite{Steph}. 

The basis of the theory of Lie symmetries lies in the invariance of differential equations under one-parameter transformations of their variables. These transformations form a local group of transformations (Lie (point) symmetries), which depend on a continuous parameter, and project any solution of the equation into another solution. Groups of transformations are fully characterized by their infinitesimal generators, which will form the corresponding Lie algebra. This method has been extensively investigated. Classical references about this subject are the textbooks \cite{Steph}, \cite{BK}, \cite{Olver}, \cite{wint}.

A standard method to find solutions of a PDE (partial differential equation) can be implemented by using Lie symmetries: each Lie symmetry leads to a similarity reduction for the PDE which allows us to reduce by one the number of independent variables. These procedures, as well as its numerous applications, have been widely studied  \cite{albares2}.

As it is well known, a PDE is considered integrable when it can be derived through the Lax equation associated to a spectral problem \cite{kono}. As mentioned, Lie symmetries for PDEs are very popular in literature, however, the identification of Lie symmetries of the associated spectral problem of an integrable system is much less frequent \cite{legare}, \cite{estevez1}. Nevertheless, the determination of the symmetries of the Lax pair has the benefit that the reduction associated to each symmetry of the Lax pair provides not only the reduction of the fields, but also the reductions of the eigenfunctions and the spectral parameter.

A different problem is the determination  of the integrability of a nonlinear partial differential equation, as well as the derivation of its Lax pair in the case of integrability. The Painlev\'e property and the singular manifold method derived from such property are extremely useful procedures in order to prove the integrability and to derive the Lax pair of an integrable system \cite{villarroel1}, \cite{diaz}, \cite{albares}, \cite{estevez2}.

This paper is devoted to the study of a multi-component non\-lin\-e\-ar \\
Schr\"o\-din\-ger (NLS) Equation in $2+1$ dimensions and its associated linear problem. In Section 2 the model  is presented. We shall prove that the system has the Painlev\'e property. The singular manifold method is successfully applied in order to derive the singular manifold equations and a three-component Lax pair.

 In Section 3  we shall apply the classical Lie method to find Lie point symmetries for the system and  the associated spectral problem. These symmetries contain many arbitrary functions that yield non-trivial commutation relations among the associated generators. These commutation relations are carefully described in Section 4. In Section 5 we shall study the reductions that arise from the symmetries, obtaining three relevant similarity reductions where the spectral parameter emerges naturally as a result of the Lie method. Finally, we close with a section of conclusions.

\section{The  Singular manifold method}
In this paper we will study a ($2+1$)-dimensional multi-component nonlinear Schr\"odinger equation 
\begin{equation}
\begin{aligned}
&i\vec\alpha_{t}+\vec\alpha_{xx}+2\,m_x\,\vec\alpha=0,\\
-&i\vec\alpha^{\dagger}_{t}+\vec\alpha^{\dagger}_{xx}+2\,m_x\,\vec\alpha^{\dagger}=0,\\
&\left( m_y+\vec\alpha\,\vec\alpha^{\dagger}\right)_x=0,
\label{sys}
\end{aligned}
\end{equation}
where $\vec\alpha(x,y,t)=\left(\begin{array}{l} \alpha_1(x,y,t) \\ \alpha_2(x,y,t)\end{array}\right)$ and $\vec\alpha^{\dagger}$ is the complex conjugate of $\vec\alpha$.  Besides that, $m(x,y,t)$ is 
a real scalar function related to the probability density $\vec\alpha \,\vec\alpha^{\dagger}$ through the third of the equations\,(\ref{sys}).

 The reduction $x=y$ of (\ref{sys}) yields the Manakov
system \cite{manakov}, which is often also called vector NLS system
\cite{apt}. Integrability properties of this Manakov system and
its Painlev\'e property are described in references \cite{lu}, \cite{wang}.
Different generalizations of this Manakov system and their solutions have been recently studied in  \cite{diaz}, \cite{albares}.

Furthermore (\ref{sys})  is the multi-component generalization of a system that has been discussed by several authors
 \cite{fokas}, \cite{chakravarty}, \cite{radha} and whose lump solutions  have been studied in \cite{villarroel1}, \cite{villarroel2}.
\subsection{Painlev\'e Property}

It is easy to check whether or not (\ref{sys}) has the Painlev\'e property. The existence of such property requires that all the solutions of  (\ref{sys}) 
are  single-valued in the initial conditions. This requirement means that the fields should be expanded as 
\begin{align}
&\alpha_1(x,y,t)=\sum_{j=0}^{\infty}a_j(x,y,t)\left[\phi(x,y,t)\right]^{j-1},&\nonumber \\
&\alpha_2(x,y,t)=\sum_{j=0}^{\infty}b_j(x,y,t)\left[\phi(x,y,t)\right]^{j-1},&\nonumber \\
&m(x,y,t)=\sum_{j=0}^{\infty}c_j(x,y,t)\left[\phi(x,y,t)\right]^{j-1},\label{2}
\end{align}
where $\phi(x,y,t)=0$ is the manifold of movable singularities. Substitution of (\ref{2}) in (\ref{sys}) 
provides five polynomials in $\phi$ whose coefficients should be $0$. 
It results in five recursion relations for the coefficients. 
The calculation can be easily performed 
with the assistance of a symbolic package such as MAPLE. We omit here the details for the benefit of the reader, but the result is that
 the recursion relations present four resonances in $j=1,3,4,6$, which are identically satisfied. Therefore we can conclude that 
(\ref{sys}) has the Painlev\'e property (PP).

The relation between integrability and Painlev\'e property is an extremely interesting  question.  The existence of  a Lax pair is usually considered as
 the best proof of the integrability of a nonlinear PDE. 
The problem of the identification of the Lax pair for a PDE which has the PP is
 a non-trivial problem which can be approached through the singular manifold method (SMM) \cite{weiss}.  The SMM is based on the truncation of equation (\ref{2})
 at the constant level. 
It means that this truncation can be considered as an auto-B\"acklund transformation of the form:
\begin{align}
&\hat{\alpha}_1=\alpha_1+A_1\,\frac{\phi_x}{\phi}, \quad\quad\quad \hat{\alpha}_1^{\dag}=\alpha_1^{\dag}+A_1^{\dag}\,\frac{\phi_x}{\phi},&\nonumber \\
&\hat{\alpha}_2=\alpha_2+A_2\,\frac{\phi_x}{\phi}, \quad\quad\quad\hat{\alpha}_2^{\dag}=\alpha_2^{\dag}+A_2^{\dag}\,\frac{\phi_x}{\phi},&\nonumber \\
&\hat{m}=m+\frac{\phi_x}{\phi},\label{3}
\end{align}
where $\left\{\alpha_1, \alpha_2, m\right\}$ are the seed fields and $\left\{\hat{\alpha}_1, \hat{\alpha}_2, \hat{m}\right\}$ the iterated ones. Besides that, $\phi$ is the singular manifold for the seed solution. Truncation means that $A_1, A_2$ and  $\phi$ should obey some equations which are known as the  singular manifold equations.
\subsection{Singular manifold equations}
Substitution of (\ref{3}) in (\ref{sys}) yields five polynomials in $\phi$. Each coefficient of these polynomials should be 0. The cumbersome calculations can be handled with MAPLE and they allow us to write the seed fields in terms of the singular manifold in the following form:
\begin{align}
&\alpha_1=-A_1\left(\frac{\left(A_1\right)_x}{A_1}+\frac{v+ir}{2}\right),\nonumber\\
&\alpha_2=-A_2\left(\frac{\left(A_2\right)_x}{A_2}+\frac{v+ir}{2}\right),\nonumber\\
&m_x=-\,\frac{1}{4}\left(v_x+\frac{v^2+r^2}{2}+\int r_t\,dx\right).\label{4}\end{align}
We have introduced for convenience the useful definitions:
\begin{equation}
v=\frac{\phi_{xx}}{\phi_x},\qquad r=\frac{\phi_t}{\phi_x}
,\qquad q=\frac{\phi_y}{\phi_x},\label{5}
\end{equation}
whose cross-derivatives trivially yield
\begin{subequations}
\begin{equation}
v_t=\left(r_x+r\,v\right)_x,\label{6a}
\end{equation}    
\begin{equation}
v_y=\left(q_x+q\,v\right)_x.\label{6b}
\end{equation}
\end{subequations}
Furthermore, the equations to be satisfied for  $A_1, A_2$ and  $\phi$ can be listed as
\begin{align}
&-i\,\frac{\left(A_1\right)_t}{A_1}=\frac{\left(A_1\right)_{xx}}{A_1}+v_x+i\,r_x+2\,m_x,\nonumber\\
&-i\,\frac{\left(A_2\right)_t}{A_2}=\frac{\left(A_2\right)_{xx}}{A_2}+v_x+i\,r_x+2\,m_x,\label{7}
\end{align}
and
\begin{align}
&q=A_1\,A_1^{\dag}+A_2A_2^{\dag},\label{8}
\end{align}
\begin{align}
\int r_y\,dx=-q\,r&+i\left[\left(A_1\right)_x\,A_1^{\dag}-A_1\,\left(A_1\right)_x^{\dag}\right]\nonumber\\
&+i\left[\left(A_2\right)_xA_2^{\dag}-A_2\,\left(A_2\right)^{\dag}_x\right].\label{9}
\end{align}

\subsection{Lax pair}
Equations (\ref{4}) can be linearized through the introduction of three-complex functions $\left\{\psi, \chi,\eta\right\}$ such that 
\begin{subequations}
\begin{align}
&A_1=\frac{\chi}{\psi},\quad\quad \quad\quad \quad
A_2=\frac{\rho}{\psi},\label{10a}\\
&v=\frac{\psi^{\dag}_x}{\psi^{\dag}}+\frac{\psi_x}{\psi},
\quad\quad\quad r=i\left(\frac{\psi^{\dag}_x}{\psi^{\dag}}-\frac{\psi_x}{\psi}\right).\label{10b}
\end{align}
\end{subequations}
\subsubsection*{Spatial part of the Lax pair}
Substitution of the definitions (\ref{10a})-(\ref{10b}) in equation (\ref{8}) yields for $q$ the expression

\begin{equation}
q=\frac{\chi\,\chi^{\dag}+\rho\,\rho^{\dag}}{\psi\,\psi^{\dag}},\label{11}
\end{equation}
and substitution of  (\ref{10a})-(\ref{10b}) in (\ref{4}) trivially results in
\begin{subequations}
\begin{equation}
\chi_x= -\,\alpha_1\,\psi\quad\Rightarrow \chi_x^{\dag}=- \,\alpha_1^{\dag}\,\psi^{\dag},
\label{12a}
\end{equation}    
\begin{equation}
\rho_x=  -\,\alpha_2\,\psi\quad\Rightarrow \rho_x^{\dag}=-\,\alpha_2^{\dag}\,\psi^{\dag}.\label{12b}
\end{equation}
\end{subequations}
Furthermore, from equations (\ref{9}) and (\ref{6b}) we have
\begin{equation}
\frac{\psi^{\dag}_y}{\psi^{\dag}}+\frac{\psi_y}{\psi}+\alpha_1\,\frac{\chi^{\dag}}{\psi^{\dag}}-\alpha_1^{\dag}\,\frac{\chi}{\psi}+\alpha_2\,\frac{\rho^{\dag}}{\psi^{\dag}}-\alpha_2^{\dag}\,\frac{\rho}{\psi}= 0,\label{13}
\end{equation}
\begin{equation}
i\left[\frac{\psi^{\dag}_y}{\psi^{\dag}}-\frac{\psi_y}{\psi}+\alpha_1\,\frac{\chi^{\dag}}{\psi^{\dag}}+\alpha_1^{\dag}\,\frac{\chi}{\psi}+\alpha_2\,\frac{\rho^{\dag}}{\psi^{\dag}}+\alpha_2^{\dag}\,\frac{\rho}{\psi}\right]=0,\label{14}
\end{equation}
which can be combined in order to obtain
\begin{equation}
\psi_y=-\,\alpha_1^{\dag}\,\chi-\,\alpha_2^{\dag}\,\rho,\quad\quad\,
\psi^{\dag}_y=-\,\alpha_1\,\chi^{\dag}-\alpha_2\,\rho^{\dag}.\label{15}
\end{equation}
Therefore, the spatial part of the Lax pair is 
\begin{equation}
\begin{pmatrix}  \partial_y\,\psi \\  \partial_x\,\chi\\  \partial_x\, \rho \end{pmatrix}=
\begin{pmatrix} &0 &-\,\alpha_1^{\dag}&-\,\alpha_2^{\dag} \\ &-\,\alpha_1&0&0\\ &-\,\alpha_2 &0&0 \end{pmatrix}\begin{pmatrix} \psi \\ \chi\\ \rho \end{pmatrix},\label{16}
\end{equation}
and its complex conjugate.

\subsubsection*{Temporal part of the Lax pair}
Substitution of (\ref{10a})-(\ref{10b}) in (\ref{4}) and (\ref{6a}) yields
\begin{align}
&\frac{\psi^{\dag}_t}{\psi^{\dag}}+\frac{\psi_t}{\psi}-i\left(\frac{\psi^{\dag}_{xx}}{\psi^{\dag}}-
\frac{\psi_{xx}}{\psi}\right)= 0,\\&
\frac{\psi^{\dag}_t}{\psi^{\dag}}-\frac{\psi_t}{\psi}-i\left(\frac{\psi^{\dag}_{xx}}{\psi^{\dag}}+
\frac{\psi_{xx}}{\psi}-4\,m_x\right)= 0,
\end{align}
which can be combined as

\begin{equation}
\psi_t=-i\,\psi_{xx}-\,2\,i\,m_x\,\psi,\quad\quad
\psi_t^{\dag}=i\,\psi^{\dag}_{xx}+2\,i\,m_x\,\psi^{\dag}.
\end{equation}
From (\ref{7}), we have
\begin{subequations}
\begin{align}
& \chi_t=-i\,\left(\alpha_1\right)_x\,\psi+i\,\alpha_1\,\psi_x,\quad\quad \chi^{\dag}_t=i\,\left(\alpha_1\right)^{\dag}_x\,\psi^{\dag}-i\,\left(\alpha_1\right)^{\dag}\,\psi^{\dag}_x,\label{20a}\\
&  \rho_t=-i\,\left(\alpha_2\right)_x\,\psi+i\,\alpha_2\,\psi_x,\quad\quad \rho^{\dag}_t=i\,\left(\alpha_2\right)^{\dag}_x\,\psi^{\dag}-
i\,\left(\alpha_2\right)^{\dag}\,\psi^{\dag}_x\label{20b}.
\end{align}
\end{subequations}
It allows us to write the temporal part of the Lax pair as 
\begin{equation}
\begin{pmatrix} \partial_t\,\psi \\  \partial_t\,\chi\\  \partial_t\,\rho \end{pmatrix}=
i\,\begin{pmatrix} &-\partial_{xx}-2\,m_x &0&0 \\ &-\partial_x\,\alpha_1+\alpha_1\,\partial_x&0&0\\ &-\partial_x \alpha_2+\alpha_2\,\partial_x &0&0 \end{pmatrix}\begin{pmatrix} \psi \\ \chi\\ \rho \end{pmatrix},\label{21}
\end{equation}
and its complex conjugate.

Equations (\ref{16}) and  
(\ref{21})  (and their complex conjugates) are a three-com\-po\-nent Lax pair for (\ref{sys}). The relation between the singular manifold  $\phi$ and the eigenfunctions $\left\{\psi, \chi,\eta\right\}$ can be easily established by combining (\ref{5}), (\ref{10b}),  and (\ref{11}) as the exact derivative:
\begin{equation}
d\phi=\psi\,\psi^{\dag}\,dx+\left(\chi\,\chi^{\dag}+\rho\,\rho^{\dag}\right)\,dy+i\,\left(\psi\,\psi^{\dag}_x-\psi^{\dag}\,
\psi_x\right)\,dt.
\end{equation}

\section{Classical Lie symmetries}
In this section, the Lie symmetry analysis is performed for the Lax pair given in equations (\ref{16}) and (\ref{21}) by applying the classical Lie method \cite{Lie}, \cite{Ovs}.

Let us consider a one-parameter Lie  group of infinitesimal transformations of the independent variables $\left\{x,y,t\right\}$, the three fields $\left\{\alpha_1,\alpha_2,m\right\}$ and the eigenfunctions $\left\{\psi, \chi,\rho\right\}$, given by:

\begin{equation}
\begin{aligned}
\tilde{x}&= x+\varepsilon\, \xi_1(x,y,t,\alpha_1,\alpha_2,m,\psi, \chi,\rho)+\mathcal{O}(\varepsilon^2),\\
\tilde{y}&= y+\varepsilon\, \xi_2(x,y,t,\alpha_1,\alpha_2,m,\psi, \chi,\rho)+\mathcal{O}(\varepsilon^2),\\
\tilde{t}&= t+\varepsilon\, \xi_3(x,y,t,\alpha_1,\alpha_2,m,\psi, \chi,\rho)+\mathcal{O}(\varepsilon^2),\\
\tilde{\alpha}_1&=\alpha_1+\varepsilon\, \eta_1(x,y,t,\alpha_1,\alpha_2,m,\psi, \chi,\rho)+\mathcal{O}(\varepsilon^2),\\
\tilde{\alpha}_2&= \alpha_2+\varepsilon\,\eta_2(x,y,t,\alpha_1,\alpha_2,m,\psi, \chi,\rho)+\mathcal{O}(\varepsilon^2),\\
\tilde{m}&= m+\varepsilon\, \eta_3(x,y,t,\alpha_1,\alpha_2,m,\psi, \chi,\rho)+\mathcal{O}(\varepsilon^2),\\
\tilde{\psi}&= \psi+\varepsilon\, \phi_1(x,y,t,\alpha_1,\alpha_2,m,\psi, \chi,\rho)+\mathcal{O}(\varepsilon^2),\\
\tilde{\chi}&= \chi+\varepsilon\, \phi_2(x,y,t,\alpha_1,\alpha_2,m,\psi, \chi,\rho)+\mathcal{O}(\varepsilon^2),\\
\tilde{\rho}&= \rho+\varepsilon\, \phi_3(x,y,t,\alpha_1,\alpha_2,m,\psi, \chi,\rho)+\mathcal{O}(\varepsilon^2),
\label{infinitesimal}
\end{aligned}
\end{equation}
where $\varepsilon$ is the group parameter and $\xi_i, \eta_i, \phi_i$, $i=1,...,3$ are the components of the related  vector field
\begin{equation}\label{X}
X = \xi_1 \frac{\partial}{\partial x} + 
\xi_2 \frac{\partial}{\partial y}+
\xi_3 \frac{\partial}{\partial t}+
\eta_1 \frac{\partial}{\partial \alpha_1}+
\eta_2 \frac{\partial}{\partial \alpha_2}+
\eta_3 \frac{\partial}{\partial m}+
\phi_1 \frac{\partial}{\partial \psi }+
\phi_2 \frac{\partial}{\partial \chi }+
\phi_3 \frac{\partial}{\partial \rho }.
\end{equation}

This infinitesimal transformation induces a well known one in the derivatives of the fields \cite{Steph}, \cite{Olver}, and it  must leave invariant the set of solutions of (\ref{16})-(\ref{21}). This procedure yields an overdetermined system of PDEs for the infinitesimals called the determining equations, whose solutions (the calculations are routinary and have been handle with MAPLE)  provide the desired Lie symmetries. The result is:
\begin{subequations}
\begin{equation}
\begin{aligned}
&\xi_1=4\,\dot{K_1}(t)\,x+2\,K_2(t),\\
&\xi_2=2\,C_1(y),\\
&\xi_3=8\,K_1(t),\\
&\eta_1=\left[i\,\left(\ddot{K_1}(t)\,x^2+\dot{K_2}(t)\,x+K_3(t)+C_2(y)\right)-2\,\dot{K_1}(t)-C_1'(y)\right]\alpha_1,\\
&\quad+\left[C_4(y)+i\,C_5(y)\right]\alpha_2,\\
&\eta_2=\left[i\,\left(\ddot{K_1}(t)\,x^2+\dot{K_2}(t)\,x+K_3(t)+C_3(y)\right)-2\,\dot{K_1}(t)-C_1'(y)\right]\alpha_2,\\
&\quad-\left[C_4(y)-i\,C_5(y)\right]\alpha_1,\\
&\eta_3=-\,4\,\dot{K_1}(t)\,m+\frac{1}{6}\dddot{K_1}(t)\,x^3+\frac{1}{4}\ddot{K_2}(t)\,x^2+\frac{1}{2}\dot{K_3}(t)\,x+\delta(y,t),\label{simEC}
\end{aligned}
\end{equation}
and
\begin{equation}
\begin{aligned}
&\phi_1=\left[-i\left(\ddot{K_1}(t)\,x^2+\dot{K_2}(t)\,x+K_3(t)\right)-2\,\dot{K_1}(t)+\lambda\right]\psi,\\
&\phi_2=\left[i\,C_2(y)-C_1'(y)+\lambda\right]\chi+\left[C_4(y)+i\,C_5(y)\right]\rho,\\
&\phi_3=\left[i\,C_3(y)-C_1'(y)+\lambda\right]\rho-\left[C_4(y)-i\,C_5(y)\right]\chi,
\label{simLP}
\end{aligned}
\end{equation}
\end{subequations}
where we have used  the convection $\,\,\dot{}=\frac{d}{dt}$ and $\,\,'=\frac{d}{dy}$.

These Lie symmetries depend on a set of nine arbitrary functions and one arbitrary constant, listed as,
\begin{itemize}
\item Three arbitrary real functions $K_j(t), \,j=1...3$, which depend exclusively on the temporal coordinate $t$.
\item Five arbitrary real functions $C_j(y), \,j=1...5$, which depend on the coordinate $y$.
\item An arbitrary real function $\delta(y,t)$.
\item Furthermore, these symmetries include an arbitrary constant $\lambda$.  We shall prove later that this constant  which will play the role of the spectral parameter in the $(1+1)$-reductions of the Lax pair. 
\end{itemize}
Lie symmetries for the Lax pair (\ref{16})-(\ref{21}) generalize, extend and include all the Lie symmetries obtained for the multi-component NLS (\ref{sys}). Symmetries given in (\ref{simEC}) can be analogously derived by implementing a similar procedure over the starting system of PDEs (\ref{sys}), whereas symmetries (\ref{simLP}) correspond to the transformation of the eigenfunctions of the Lax pair. It is also worthwhile to remark that the only additional symmetry that corresponds strictly to the Lax pair itself is the one associated with the arbitrary constant $\lambda$.

\section{Commutation relations}

In this section, we will analyze the commutation relations among the infinitesimal generators associated to the Lie symmetries for the Lax pair obtained in the previous section.

The infinitesimal generators associated to each symmetry are listed below.
\begin{itemize}

\item We shall denote as $X^{[j]}_{\{K_j(t)\}}, \, {j=1...3}$, the generators associated to the three arbitrary functions of $t$. 

\item Let  $Y^{[l]}_{\{C_l(y)\}}, \,{l=1...5}$, be the generators associated to the arbitrary functions of $y$.  
\item The generators associated to the arbitrary function $\delta(y,t)$ is denoted as $Z_{\{\delta(y,t)\}}$. 

\item Finally, we have defined as $\Lambda_{\{\lambda\}}$ the infinitesimal generator related to the arbitrary constant $\lambda$.
\end{itemize}
With this notation, we have the following ten generators:

\begin{equation}
\begin{aligned}
X_{{\{K_1(t)\}}}^{[1]}&=\frac{1}{6}\,x^3\,\dddot{K_1}\frac{\partial}{\partial m}+i\,x^2\,\ddot{K_1}\left(\alpha_1\,\frac{\partial}{\partial \alpha_1}+\alpha_2\,\frac{\partial}{\partial \alpha_2}-\psi\,\frac{\partial}{\partial \psi}\right)\\
&+2\dot{K_1}\left(2\,x\,\frac{\partial}{\partial x}-\alpha_1\,\frac{\partial}{\partial \alpha_1}-\alpha_2\,\frac{\partial}{\partial \alpha_2}-2\,m\,\frac{\partial}{\partial m}-\psi\,\frac{\partial}{\partial \psi}\right) +8\,K_1\,\frac{\partial}{\partial t},\\
X^{[2]}_{\{K_2(t)\}}&=\frac{1}{4}\,x^2\,\ddot{K_2}\,\frac{\partial}{\partial m}+i\,x\,\dot{K_2}\,\left(\alpha_1\,\frac{\partial}{\partial \alpha_1}+\alpha_2\,\frac{\partial}{\partial \alpha_2}-\psi\,\frac{\partial}{\partial \psi}\right)+2\,K_2\,\frac{\partial}{\partial x},\\
X^{[3]}_{\{K_3(t)\}}&=\frac{1}{2}\,x\,\dot{K_3}\,\frac{\partial}{\partial m}+i\,K_3\,\left(\alpha_1\,\frac{\partial}{\partial \alpha_1}+\alpha_2\,\frac{\partial}{\partial \alpha_2}-\psi\,\frac{\partial}{\partial \psi}\right),\\
Y_{\{C_1(y)\}}^{[1]}&=-\,C_1'\left(\alpha_1\,\frac{\partial}{\partial \alpha_1}+\alpha_2\,\frac{\partial}{\partial \alpha_2}+\chi\,\frac{\partial}{\partial \chi}+\rho\,\frac{\partial}{\partial \rho}\right)+2\,C_1\,\frac{\partial}{\partial y},\\
Y^{[2]}_{\{C_2(y)\}}&=i\,C_2\left(\alpha_1\,\frac{\partial}{\partial \alpha_1}+\chi\,\frac{\partial}{\partial \chi}\right),\\
Y^{[3]}_{\{C_3(y)\}}&=i\,C_3\left(\alpha_2\,\frac{\partial}{\partial \alpha_2}+\rho\,\frac{\partial}{\partial \rho}\right),\\
Y^{[4]}_{\{C_4(y)\}}&=C_4\left(\alpha_2\,\frac{\partial}{\partial \alpha_1}-\alpha_1\,\frac{\partial}{\partial \alpha_2}+\rho\,\frac{\partial}{\partial \chi}-\chi\,\frac{\partial}{\partial \rho}\right),\\
Y^{[5]}_{\{C_5(y)\}}
&=i\,C_5\left(\alpha_2\,\frac{\partial}{\partial \alpha_1}+\alpha_1\,\frac{\partial}{\partial \alpha_2}+\rho\,\frac{\partial}{\partial \chi}+
\chi\,\frac{\partial}{\partial \rho}\right),\\
Z_ {\{\delta(y,t)\}}&=\delta\,\frac{\partial}{\partial m},\\
\Lambda_{\{\lambda\}}&=\left(\psi\,\frac{\partial}{\partial \psi}+\chi\,\frac{\partial}{\partial \chi}+\rho\,\frac{\partial}{\partial \rho}\right).
\end{aligned}
\end{equation}

According to \cite{Steph}, symmetry generators of PDEs can be classified in two classes, the ones associated to the arbitrary constants and the ones associated to the arbitrary functions. While those of the first type will give rise to a Lie algebra, the infinitesimal generators that depend on arbitrary functions do not, since we are dealing with an infinite-dimensional basis of generators. Nonetheless, it can be proved that the commutator of two symmetry generators is also a generator of a symmetry.

Commutations relations among these operators may be performed. 
The convention used is that each $\{j,l\}$-element of the table corresponds to the operation $\left[\mathcal{X}^{[j]}_{\{\mathcal{\kappa}_j\}},\mathcal{X}^{[l]}_{\{\mathcal{\kappa}_l\}}\right]$, where $\mathcal{X}^{[j]}_{\{\mathcal{\kappa}_j\}}$ is the generator associated to a function $\mathcal{\kappa}_j$ of its characteristic independent variables. The results are presented in two different tables for greater usability for the reader, but they should not be interpreted separately:

\begin{table}[H]
\centering
\resizebox{\textwidth}{!} {
\begin{tabular}{c||c|c|c|c|c}
  & $
  X^{[1]}_{\{K_1\}}$ & $X^{[2]}_{\{K_2\}}$ & $X^{[3]}_{\{K_3\}}$ & $Z_{\{\delta\}}$\\
	\hline \hline
	$X^{[1]}_{\{H_1\}}$ & $X^{[1]}_{\{8H_1\dot{K_1}-8K_1\dot{H_1}\}}$  & $X^{[2]}_{\{8H_1\dot{K_2}-4K_2\dot{H_1}\}}$ & $X^{[3]}_{\{8H_1\dot{K_3}\}}$ & $Z_{\{8H_1\partial_t(\delta)+4\delta\dot{H_1}\}}$\\ \hline
	$X^{[2]}_{\{H_2\}}$ & $-X^{[2]}_{\{8K_1\dot{H_2}-4H_2\dot{K_1}\}}$ & $X^{[3]}_{\{2H_2\dot K_2-2K_2\dot H_2\}}$ & $Z_{\{H_2\dot{K_3}\}}$ & 0\\ \hline
	$X^{[3]}_{\{H_3\}}$ & $-X^{[3]}_{\{8K_1\dot{H_3}\}}$ & $-Z_{\{K_2\dot{H_3}\}}$ & 0 & 0\\ \hline
    $Z_{\{\gamma\}}$ & $-Z_{\{8K_1\partial_t(\gamma)+4\gamma\dot{K_1}\}}$ & 0 & 0 & 0 \\ \hline
    \end{tabular}}
\end{table}

\begin{table}[H]
\centering
\resizebox{\textwidth}{!} {
\begin{tabular}{c||c|c|c|c|c|c|c}
  & $Z_{\{\delta\}}$ & $Y^{[1]}_{\{C_1\}}$ & $Y^{[2]}_{\{C_2\}}$ & $Y^{[3]}_{\{C_3\}}$ & $Y^{[4]}_{\{C_4\}}$ & $Y^{[5]}_{\{C_5\}}$\\
	\hline \hline
	$Z_{\{\gamma\}}$ & $0$ & $-Z_{\left\{2\,C_1\partial_y(\gamma)\right\}}$ & 0 & 0 & 0 & 0 \\ \hline
	$Y^{[1]}_{\{J_1\}}$ & $Z_{\left\{2\,J_1\partial_y(\delta)\right\}}$ & $Y^{[1]}_{\{2J_1\,C_1'-2C_1\,J_1'\}}$ & $Y^{[2]}_{\{2J_1C_2'\}}$ & $Y^{[3]}_{\{2J_1C_3'\}}$ & $Y^{[4]}_{\{2J_1C_4'\}}$ & $Y^{[5]}_{\{2J_1C_5'\}}$\\ \hline
	$Y^{[2]}_{\{J_2\}}$ & 0 & $-Y^{[2]}_{\{2C_1J_2'\}}$ & 0 & 0 & $-Y^{[5]}_{\{J_2\,C_4\}}$ & $Y^{[4]}_{\{J_2\,C_5\}}$\\ \hline
    $Y^{[3]}_{\{J_3\}}$ & 0 & $-Y^{[3]}_{\{2C_1J_3'\}}$ & 0 & 0 & $Y^{[5]}_{\{J_3\,C_4\}}$ & $-Y^{[4]}_{\{J_3\,C_5\}}$\\ \hline
    $Y^{[4]}_{\{J_4\}}$ & 0 & $-Y^{[4]}_{\{2C_1J_4'\}}$ & $Y^{[5]}_{\{C_2\,J_4\}}$ & $-Y^{[5]}_{\{C_3\,J_4\}}$ & 0 & $-Y^{[2]}_{\{2\,J_4C_5\}}+Y^{[3]}_{\{2\,J_4C_5\}}$\\ \hline
    $Y^{[5]}_{\{J_5\}}$ & 0 & $-Y^{[5]}_{\{2C_1J_5'\}}$ & $-Y^{[4]}_{\{C_2\,J_5\}}$ & $Y^{[4]}_{\{C_3\,J_5\}}$ & $Y^{[2]}_{\{2\,C_4J_5\}}-Y^{[3]}_{\{2\,C_4J_5\}}$ & 0\\
\hline	\end{tabular}
    }
\end{table}

Notice that  the generator $\Lambda_{\{\lambda\}}=\left(\psi\,\partial_{\psi}+\chi\,\partial_{ \chi}+\rho\,\partial_{\rho}\right)$ commutes with all other generators. Besides that, $\left[X^{[j]}_{\{K_j(t)\}},Y^{[l]}_{\{C_l(y)\}}\right]=0$ for every value of $j=1...3, l = 1...5$. Indeed, it may be observed that every commutator of two infinitesimal generators provides a non-trivial result, due to the presence of arbitrary functions \cite{Olver}.

We should remark that, in general, these infinitesimal generators do not form a Lie algebra, but it is possible to construct a finite-dimensional Lie algebra by selecting special values for the arbitrary functions. Some relevant works about this topic have been developed in \cite{Champ87}, \cite{David85}, where Kac-Moody type algebras have been obtained through a polynomial dependence for the arbitrary functions.  

\section{Similarity reductions}\label{reductions}

Similarity reductions may be achieved by solving the characteristic system 
\begin{equation} \label{characteristic}
\frac{dx}{\xi_{1}}=\frac{dy}{\xi_{2}}=\frac{dt}{\xi_{3}}=\frac{d\alpha_1}{\eta_{1}}=\frac{d\alpha_2}{\eta_{2}}=\frac{dm}{\eta_{3}}=\frac{d\psi}{\phi_{1}}=\frac{d\chi}{\phi_{2}}=\frac{d\rho}{\phi_{3}}.
\end{equation}

In the following, we shall summarize the notation used for the reduced variables, reduced fields and reduced eigenfunctions:

\begin{equation}
\left\{
\begin{array}{ccc}
 x,y,t \rightarrow p,\,q, & &\\
\alpha_1(x,y,t) \rightarrow F(p,q), & \alpha_2(x,y,t)\rightarrow H(p,q), &  m(x,y,t)\rightarrow N(p,q),\\
\psi(x,y,t)\rightarrow \Phi(p,q), & \chi(x,y,t)\rightarrow \Sigma (p,q), &\rho(x,y,t)\rightarrow \Omega(p,q).
\end{array}
\right.
\end{equation}

The symmetries that will yield non-trivial reductions are those related to the arbitrary functions $K_1(t)$, $K_2(t)$ and $C_1(y)$, present in the transformations of the independent variables. The rest of the symmetries provide trivial reductions. Several reductions may emerge for different values of $K_1, K_2, C_1$, raising three independent reductions.

We will introduce the following shorthand notation, which will be very useful for the next calculations:

\begin{equation}I_0(t)=\frac{1}{4}\int{\frac{K_2(t)}{K_1(t)^{\frac{3}{2}}}\,dt},\quad I_1(t)=\frac{1}{4}\int{\frac{K_2(t)^2}{K_1(t)^2}\,dt},\quad I_2(t)=\frac{1}{512}\int{\frac{K_2(t)^3}{K_1(t)^{\frac{5}{2}}}\,dt}.\end{equation}

\subsection{$K_1(t)\neq 0$, $K_2(t)\neq 0$, $C_1(y)\neq 0$}

By solving the characteristic system (\ref{characteristic}), the following results have been obtained
\begin{itemize}
\item Reduced variables
\begin{equation}
p=\frac{x}{K_1(t)^{\frac{1}{2}}}-I_0(t), \qquad q=4\int{\frac{dy}{C_1(y)}}-\int{\frac{dt}{K_1(t)}}.
\end{equation}
\item Reduced fields
\begin{equation}
\begin{aligned}
\alpha_1(x,y,t)&=\frac{2\,F(p,q)}{K_1(t)^{\frac{1}{4}}\,C_1(y)^{\frac{1}{2}}\,}\,e^{\left\{\frac{i}{8}\left[\frac{\dot{K}_1(t)}{K_1(t)}x^2+\frac{K_2(t)}{K_1(t)}x-I_1(t)\right]\right\}},\\
\alpha_2(x,y,t)&=\frac{2\,H(p,q)}{K_1(t)^{\frac{1}{4}}\,C_1(y)^{\frac{1}{2}}\,}\,e^{\left\{\frac{i}{8}\left[\frac{\dot{K}_1(t)}{K_1(t)}x^2+\frac{K_2(t)}{K_1(t)}x-I_1(t)\right]\right\}},\\
\small m(x,y,t)&=\frac{x^3}{24\,K_1(t)^{\frac{1}{2}}}\,\left[K_1(t)^{\frac{1}{2}}\right]_{tt}+\frac{x^2}{32\, K_1(t)^{\frac{1}{2}}}\,\left[\frac{K_2(t)}{K_1(t)^{\frac{1}{2}}}\right]_t\\ &-\frac{x}{32}\,\dot{I}_1(t)+\frac{N(p,q)+I_2(t)}{K_1(t)^{\frac{1}{2}}},\,\normalsize
\end{aligned}
\end{equation}
where the subscript $(\cdot)_t$ denotes the derivative with respect to the coordinate $t$.
\item Reduced eigenfunctions
\begin{equation}
\begin{aligned}
\psi(x,y,t)&=\frac{\Phi(p,q)}{2\,K_1(t)^{\frac{1}{4}}}\,e^{\left\{-\frac{i}{8}\left[\frac{\dot{K}_1(t)}{K_1(t)}\,x^2+\frac{K_2(t)}{K_1(t)}\,x-I_1(t)\right]+\frac{\lambda}{8}\int{\frac{dt}{K_1(t)}}\right\}},\\
\chi(x,y,t)&=\frac{\Sigma(p,q)}{C_1(y)^{\frac{1}{2}}}\,e^{\left\{\frac{\lambda}{8}\int{\frac{dt}{K_1(t)}}\right\}},\\\rho(x,y,t)&=\frac{\Omega(p,q)}{C_1(y)^{\frac{1}{2}}}\,e^{\left\{\frac{\lambda}{8}\int{\frac{dt}{K_1(t)}}\right\}}.
\end{aligned}
\end{equation}
\item Reduced spectral problem

By substituting the reductions in the ($2+1$)-Lax pair (\ref{16})-(\ref{21}) we obtain the following ($1+1$)-Lax pair:
\begin{eqnarray}
&&\Phi_{pp}+\left(2\,N_p-\frac{i}{8}\,\lambda\right)\Phi-i\,F^{\dagger}\,\Sigma-i\,H^{\dagger}\,\Omega=0,\nonumber\\
&&\Sigma_p+F\,\Phi=0,\label{I-S}\\
&&\Omega_p+H\,\Phi=0,\nonumber\\
 &&\nonumber\\
&&\Phi_{q}+F^{\dagger}\,\Sigma+H^{\dagger}\,\Omega=0,\nonumber\\
&&\Sigma_q+i\left(F\,\Phi_p-F_p\,\Phi\right)-\frac{\lambda}{8}\,\Sigma=0,\label{I-T}\\
&&\Omega_q+i\left(H\,\Phi_p-H_p\,\Phi\right)-\frac{\lambda}{8}\,\Omega=0.\nonumber
\end{eqnarray}
\item Reduced Equations

The compatibility condition between (\ref{I-S})-(\ref{I-T}) will provide the reduced equations (and its complex conjugate)
\begin{equation}
\begin{aligned}
&iF_q-F_{pp}-2FN_p=0,\\
&iH_q-H_{pp}-2HN_p=0,\\
&\left(N_q+FF^{\dagger}+HH^{\dagger}\right)_p=0,
\end{aligned}
\end{equation}
which prove to be a nonlocal multi-component NLS Equation in $1+1$ dimensions, expressed for the complex conjugate fields $\{F^{\dagger}\,,H^{\dagger}\}$ with density of probability $N_q$. This reduction corresponds to the Manakov system \cite{manakov}, \cite{apt}.
\end{itemize}
We may remark that the same reductions for the Lax pair and consequently, for the equations, will be obtained by performing the similarity reductions for the case with $K_1(t)\neq 0$, $C_1(y)\neq 0$,  $K_2(t)=0$, although the reductions for the independent variables, fields and eigenfunctions are different.

\subsection{$K_1(t)\neq 0$, $K_2(t)\neq 0$, $C_1(y)= 0$}
Integration of (\ref{characteristic}) provides the following results
\begin{itemize}
\item Reduced variables
\begin{equation}
p=\frac{x}{K_1(t)^{\frac{1}{2}}}-I_0(t), \qquad q=y.
\end{equation}
\item Reduced fields
\begin{equation}
\begin{aligned}
\alpha_1(x,y,t)&=\frac{F(p,q)}{K_1(t)^{\frac{1}{4}}}\,e^{\left\{\frac{i}{8}\left[\frac{\dot{K}_1(t)}{K_1(t)}\,x^2+\frac{K_2(t)}{K_1(t)}\,x-I_1(t)\right]\right\}},\\
\alpha_2(x,y,t)&=\frac{H(p,q)}{K_1(t)^{\frac{1}{4}}}\,e^{\left\{\frac{i}{8}\left[\frac{\dot{K}_1(t)}{K_1(t)}\,x^2+\frac{K_2(t)}{K_1(t)}\,x-I_1(t)\right]\right\}},\\
\small m(x,y,t)&=\frac{x^3}{24\,K_1(t)^{\frac{1}{2}}}\left[K_1(t)^{\frac{1}{2}}\right]_{tt}+\frac{x^2}{32\, K_1(t)^{\frac{1}{2}}}\,\left[\frac{K_2(t)}{K_1(t)^{\frac{1}{2}}}\right]_t\\ &-\frac{x}{32}\,\dot{I}_1(t)+\frac{N(p,q)+I_2(t)}{K_1(t)^{\frac{1}{2}}}.\normalsize
\end{aligned}
\end{equation}
\item Reduced eigenfunctions
\begin{equation}
\begin{aligned}
\psi(x,y,t)&=\frac{\Phi(p,q)}{K_1(t)^{\frac{1}{4}}}\,e^{\left\{-\frac{i}{8}\left[\frac{\dot{K}_1(t)}{K_1(t)}\,x^2+\frac{K_2(t)}{K_1(t)}\,x-I_1(t)\right]+\frac{\lambda}{8}\int{\frac{dt}{K_1(t)}}\right\}},\\
\chi(x,y,t)&=\Sigma(p,q)\,e^{\left\{\frac{\lambda}{8}\int{\frac{dt}{K_1(t)}}\right\}},\\\rho(x,y,t)&=\Omega(p,q)\,e^{\left\{\frac{\lambda}{8}\int{\frac{dt}{K_1(t)}}\right\}}.
\end{aligned}
\end{equation}
\item Reduced spectral problem
\begin{eqnarray}
&&\Phi_{pp}+\left(2N_p-\frac{i}{8}\lambda\right)\Phi=0,\nonumber\\
&&\Sigma_p+F\,\Phi=0,\label{II-S}\\
&&\Omega_p+H\,\Phi=0,\nonumber\\&&\nonumber\\
&&\Phi_{q}+F^{\dagger}\,\Sigma+H^{\dagger}\,\Omega=0.\nonumber\\
&&\lambda\,\Sigma-8\,i\left(F\,\Phi_p-F_p\,\Phi\right)=0,\label{II-T}\\
&&\lambda\,\Omega-8\,i\left(H\,\Phi_p-H_p\,\Phi\right)=0.\nonumber
\end{eqnarray}

The previous system of PDEs can be expressed equivalently to the following scalar Lax pair in $1+1$ dimensions
\begin{equation}
\begin{aligned}
&\Phi_{pp}+\left(2\,N_p-\frac{i}{8}\,\lambda\right)\Phi=0,\\
&\lambda\,\Phi_{q}-8\,i\left[\left(F^{\dagger}\,F_p+H^{\dagger}\,H_p\right)\,\Phi+N_q\,\Phi_p\right]=0.\label{II}
\end{aligned}
\end{equation}

\item Reduced Equations

The compatibility condition between (\ref{II}) yield the reduced equations (and its complex conjugate)
\begin{equation}
\begin{aligned}
&F_{pp}+2FN_p=0,\\
&H_{pp}+2HN_p=0,\\
&\left(N_q+FF^{\dagger}+HH^{\dagger}\right)_p=0.
\end{aligned}
\end{equation}
\end{itemize}

\subsection{$K_2(t)\neq 0$, $C_1(y)\neq 0$, $K_1(t)=0$}
The following reductions arise from the integration of (\ref{characteristic}),
\begin{itemize}
\item Reduced variables
\begin{equation}
p=\frac{x}{K_2(t)}-\int{\frac{dy}{C_1(y)}}, \quad q=\int{\frac{dt}{K_2(t)^2}}.
\end{equation}
\item Reduced fields
\begin{equation}
\begin{aligned}
\alpha_1(x,y,t)&=\frac{F(p,q)}{K_2(t)^{\frac{1}{2}}\,C_1(y)^{\frac{1}{2}}\,}\,e^{\left\{\frac{i}{4}\left[\frac{\dot{K}_2(t)}{K_2(t)}\,x^2+2\,p-q\right]\right\}},\\
\alpha_2(x,y,t)&=\frac{H(p,q)}{K_2(t)^{\frac{1}{2}}\,C_1(y)^{\frac{1}{2}}\,}\,e^{\left\{\frac{i}{4}\left[\frac{\dot{K}_2(t)}{K_2(t)}\,x^2+2\,p-q\right]\right\}},\\
m(x,y,t)&=\frac{x^3}{24}\,\frac{\ddot{K}_2(t)}{K_2(t)}+\frac{N(p,q)}{K_2(t)}.
\end{aligned}
\end{equation}
\item Reduced eigenfunctions
\begin{equation}
\begin{aligned}
\psi(x,y,t)&=\frac{\Phi(p,q)}{K_2(t)^{\frac{1}{2}}}\,e^{\left\{-\frac{i}{4}\left[\frac{\dot{K}_2(t)}{K_2(t)}\,x^2-\,q\right]+\frac{\lambda}{2}\int{\frac{dy}{C_1(y)}}\right\}},\\
\chi(x,y,t)&=\frac{\Sigma(p,q)}{C_1(y)^{\frac{1}{2}}}\,e^{\left\{\frac{\lambda}{2}\int{\frac{dy}{C_1(y)}}+\,i\,\frac{p}{2}\right\}},\\ \rho(x,y,t)&=\frac{\Omega(p,q)}{C_1(y)^{\frac{1}{2}}}\,e^{\left\{\frac{\lambda}{2}\int{\frac{dy}{C_1(y)}}+\,i\,\frac{p}{2}\right\}}.
\end{aligned}
\end{equation}
\item Reduced spectral problem
\begin{eqnarray}
&&\Phi_{p}-\left(F^{\dagger}\,\Sigma+H^{\dagger}\,\Omega\right)-\frac{\lambda}{2}\,\Phi=0,\nonumber\\
&&\Sigma_p+F\,\Phi+\frac{i}{2}\,\Sigma=0,\label{III-S}\\
&&\Omega_p+H\,\Phi+\frac{i}{2}\,\Omega=0,\nonumber\\
&&\nonumber\\
&&\Phi_{q}+\left(i\,H^{\dagger}_p+\frac{i\,\lambda+1}{2}\,H^{\dagger}\right)\Omega+\left(i\,F^{\dagger}_p+\frac{i\,\lambda+1}{2}\,F^{\dagger}\right)\Sigma\nonumber\\ &&\quad\quad -i\left(FF^{\dagger}+HH^{\dagger}-2\,N_p-\,\frac{\lambda^2+1}{4}\right)\Phi=0,\nonumber\\
&&\Sigma_q-\left(\frac{i\,\lambda+1}{2}\,F-i\,F_p\right)\Phi-i\,FF^{\dagger}\,\Sigma-i\,FH^{\dagger}\,\Omega=0,\label{III-T}\\
&&\Omega_q-\left(\frac{i\,\lambda+1}{2}\,H-i\,H_p\right)\Phi-i\,HF^{\dagger}\,\Sigma-i\,HH^{\dagger}\,\Omega=0.\nonumber
\end{eqnarray}
\item Reduced Equations

The compatibility condition between (\ref{III-S})-(\ref{III-T}) will provide the following system of PDEs
\begin{equation}
\begin{aligned}
&i\,F_q+\left(F_p+i\,F\right)_p+2\,FN_p=0,\\
&i\,H_q+\left(H_p+i\,H\right)_p+2\,HN_p=0,\\
&\left(N_p-FF^{\dagger}-HH^{\dagger}\right)_p=0.
\end{aligned}
\end{equation}
\end{itemize}

\section{Conclusions}

In this paper a multi-component Nonlinear Schr\"odinger Equation in $2+1$ dimensions has been presented. This system constitutes a generalization of the Manakov system to higher dimensions. The Painlev\'e test has been proved to be a powerful technique to identify the integrability of this model. Furthermore, the SMM has enabled us to derive a non-trivial three-component Lax pair for this system.

We have determined the classical Lie symmetries for a multi-component Nonlinear Schr\"{o}dinger equation in $2+1$ dimensions and its three-component Lax pair. This procedure allows us to get the infinitesimals related to the independent variables
and fields, along with those associated to the eigenfunctions. The resulting symmetries  include nine arbitrary functions of the independent variables and a single arbitrary constant, which plays the role of the spectral parameter when the spectral problem is reduced to $1+1$ dimensions. 

The commutation relations among the generators associated to each symmetry have been widely analyzed. Although the set of symmetries does not form a Lie algebra (due to the presence of arbitrary functions), these relations are consistent and closed. Eventually, we could define the Lie algebra associated to particular selections for the arbitrary functions.   

Three non-trivial reductions to $1+1$ dimensions have been derived. The reduced equations and the reduced spectral problem have been simultaneously obtained. It is important to notice that the spectral parameter arises naturally in the process of constructing the reductions, due to the symmetry procedure itself.

\section*{Acknowledgements}
 This research
has been supported by MINECO (Grant MAT2016-75955) and
Junta de Castilla y Le\'on (Grant SA045U16). P. Albares also acknowledges a predoctoral grant supported by Junta de Castilla y Le\'on.
We wish also thank Jose M. Cerver\'o for his continuous advise and helpful comments.

\end{document}